\documentclass[prd,twocolumn,showpacs,amsmath,amssymb,nofootinbib,preprintnumbers]{revtex4}
\usepackage{axodraw}
\usepackage{subfigure}
\usepackage{float}
\usepackage{wrapfig}
\usepackage{array}
\usepackage{graphics}
\usepackage{graphicx}

\newcommand{\Tr}{\text{Tr}}

\begin{document}

$\text{}$ \preprint{HD-THEP-03-36}

\title{First Order Chiral Phase Transition from a Six-fermion ''Instanton''-Interaction }

\author{Joerg Jaeckel}
\email{Jaeckel@thphys.uni-heidelberg.de}
\author{Christof Wetterich}%
 \email{C.Wetterich@thphys.uni-heidelberg.de}
\affiliation{Institut f\"ur Theoretische Physik, Universit\"at
Heidelberg, Philosophenweg 16, 69120 Heidelberg}

\begin{abstract}
We compute the first order chiral phase transition for an instanton motivated quark model
with a local  six-quark interaction. In order to compare different solutions of the gap equation
we compute the bosonic effective action -- a two particle irreducible free energy functional.
We find that the first order transition ends for a critical current quark mass, with continuous
crossover for larger quark masses. Furthermore, we investigate different possible order parameters,
including a color octet condensate. We also compare our formalism with mean field theory.
\end{abstract}
\maketitle
\section{Introduction}
The understanding of strongly interacting fermion systems is one of the great
challenges for theoretical physics, ranging from condensed matter systems and ultracold atoms
to strong interactions in particle physics. Non-perturbative methods need to be explored
in order to deal with strong effective interactions, collective phenomena, condensates
in competing channels and different characteristic physics at
``microscopic'' and ``macroscopic'' distances. In this note
we deal with the bosonic effective action based on a two-particle irreducible formalism and
the Schwinger-Dyson equations in a situation where the dominant interactions involve
more than four fermions. We also compare with mean field theory.

Indeed, for some physical systems the higher order fermion interactions play a crucial role.
As a concrete example we discuss the instanton interaction in quantum chromodynamics (QCD) with three
flavors of quarks. The violation of the axial $U(1)_{A}$-symmetry by instanton effects induces an
effective six-quark-interaction which involves all three flavors \cite{'tHooft:fv}. It has been
speculated that this instanton mediated interaction may become dominant at a characteristic momentum scale
below $1\textrm{GeV}$. We explore here a model where the instanton mediated interaction
dominates the effective theory at some given scale $\Lambda$. This scale will
act as an ultraviolet cutoff for the fluctuations with momenta $q^2<\Lambda^2$ which induce
condensates and spontaneous chiral symmetry breaking.
We concentrate on the pointlike limit of this interaction and neglect all other effective interactions
between the quarks. We do not believe that this model is a sufficient description for real QCD but
it points to some characteristic features and serves as an interesting demonstration for the more
formal part of this work.

First investigations of strongly interacting fermionic systems are often based on
Mean Field Theory (MFT) or
lowest order Schwinger-Dyson equations (SDE) \linebreak[4] \cite{Dyson:1949ha,Schwinger:1951ex}
\footnote{For some reviews on SDE's and more sophisticated approximations
see e.g. \cite{Roberts:dr,Roberts:2000aa,Alkofer:2000wg}.}.
For example, the recent studies of color superconductivity
\cite{Rapp:1997zu,Bailin:bm,Alford:1997zt,Alford:1998mk,Berges:1998rc,Schafer:1998ef,Oertel:2002pj}
are mainly based on one of these methods.
Both methods allow for a computation of the order parameter in systems which
exhibit spontaneous symmetry breaking (SSB).
However, while the SDE approach leads directly to the gap equation, the MFT approach
provides naturally a free energy functional for the bosonic composite
degrees of freedom introduced by partial bosonization via a Hubbard-Stratonovich
transformation \cite{stratonovich,Hubbard:1959ub}. The minima of the free energy are determined by the
field equation which
corresponds to a type of gap equation.
Knowledge of the free energy functional becomes necessary if the gap (or field) equation
allows solutions with different order parameters and the free energy for the different solutions
has to be compared.

Within the SDE approach the reconstruction of the free energy functional from
the gap equation is not trivial since the information about the minima (as expressed by the gap equation)
may be insufficient in order to find the whole function.
For example, it is not sure that the method used in \cite{Bowers:2002xr}
for the case of color superconductivity always works.
From this point of view MFT seems superior to SDE since it directly provides a free energy. Unfortunately, MFT
has also a severe disadvantage: partial bosonization is not unique and the results of the
MFT calculation depend strongly on the choice of the mean field (see \cite{Baier:2000yc} for
the example of the Hubbard model where this has drastic consequences). Since partial bosonization
is an exact procedure physical results should be independent of this choice.
Indeed, the ambiguity
is due to the approximation used in MFT (only fermionic fluctuations are taken into account)
and is cured by more sophisticated approximations \cite{Jaeckel:2003kf}.
Being a purely fermionic formulation SDE's do not suffer from such an ambiguity\footnote{Moreover
SDE's are one-loop exact even in the lowest approximation, while for MFT this is not
necessarily the case. This is
directly related to the ambiguity.}.

Hence, we want to find a functional which has the SDE
as its equation of motion, and which can be interpreted as a free energy.
For four fermion interactions such a functional can be related to the two particle
irreducible (2PI) effective action
\cite{Baym:1961a,Baym:1961b,Cornwall:vz,Wetterich:2002ky}.
The 2PI effective action is a functional of
fields and propagators $\Gamma^{(2PI)}[\psi,G]$. However,
for a purely fermionic system, all the information
is already contained in $\Gamma^{(2PI)}[0,G]$ where the ``fermionic background field''
is set to zero. Hence
$\Gamma^{(2PI)}[0,G]$ depends only on the bosonic variable $G$,
and therefore we will call it the Bosonic Effective Action (BEA)
\cite{Wetterich:2002ky}.
In a suitable version the BEA is a free energy functional and the condition for its local
minima precisely corresponds to the gap equation.

The work of \cite{Wetterich:2002ky} has concentrated on effective four fermion interactions. In the
present paper we want to deal with more general multi-fermion interactions.
An example of such a higher order interaction is the six-fermion interaction generated by instantons
in the case of three flavors and three colors \cite{Wetterich:2000ky,Callan:1977gz,Schafer:1996wv,Shuryak:1982hk,Shifman:uw,'tHooft:fv}.
This interaction is $U(1)_{A}$-anomalous and solves the famous $U(1)_{A}$-problem \cite{'tHooft:1986nc} in QCD.
In the simpler case of two flavors the instantons mediate a four fermion interaction which has already been
studied extensively for
chiral symmetry breaking \cite{Diakonov:vw,Diakonov:1985eg,Carter:1999xb} and was considered in the
early works on color superconductivity e.g. \cite{Berges:1998rc,Alford:1997zt}.
This investigation is generalized here to three light flavors of quarks which are perhaps
closer to realistic QCD. We also perform a more systematic
discussion of the free energy functional. We observe that the
effective interaction generated by the instantons does not only lead to interactions between color
singlet effective quark-antiquark degrees of freedom which are
finally associated to the usual spontaneous chiral symmetry breaking. It
also produces effective interactions between color octets. In a model that
goes beyond the six quark interactions the instanton effects may lead to the possibility
of octet condensation and spontaneous ``color symmetry
breaking'' \cite{Wetterich:2000ky,Wetterich:2000pp,Wetterich:1999vd}.

The paper is organized as follows. In sects. \ref{sec::bea} and \ref{sec::1vertex}
we introduce the 2PI formalism in the language of the Bosonic Effective Action (BEA).
In sect. \ref{sec::mft}
we make a short comparison to MFT and hint to some possible improvements.
We explicitly calculate the BEA for the six-fermion instanton interaction
in sect. \ref{sec::instanton1} and use
it to study chiral symmetry breaking. Finally, in sect \ref{sec::instanton2} we
investigate color octet condensates. Sect. \ref{sec::conclusions} summarizes our results and conclusions.

\section{Bosonic Effective Action}\label{sec::bea}
In this section we briefly summarize the Schwinger Dyson equations \cite{Dyson:1949ha,Schwinger:1951ex}
and the fermionic and bosonic effective action \cite{Wetterich:2002ky}
and generalize it to multi-fermion interactions. In order to
simplify the presentation we summarize all indices of the
fermionic field in $\tilde{\psi}_{\alpha}$. Here the index $\alpha$ contains
all internal indices (spin, color, flavor etc.) as well as
position or momentum. Furthermore
it also differentiates between $\psi$ and $\bar{\psi}$.
In these conventions the partition function reads
\begin{equation}
\label{equ::partitionfunction}
Z[\eta,j]=\int{\mathcal{D}}\tilde{\psi}\exp(\eta_{\alpha}\tilde{\psi}_{\alpha}
+\frac{1}{2}j_{\alpha\beta}\tilde{\psi}_{\alpha}\tilde{\psi}_{\beta}-S_{\text{int}}[\tilde{\psi}]).
\end{equation}
All terms quadratic in $\psi$ are associated to bosonic sources $j_{\alpha\beta}$
and we investigate multifermion interactions ($n$ even, $n\geq 4$)
\begin{equation}
\label{equ::iniaction}
S_{\text{int}}[\tilde{\psi}]=\sum_{n}\frac{1}{n!}\lambda^{(n)}_{\alpha_{1}...\alpha_{n}}
\tilde{\psi}_{\alpha_{1}}\cdots\tilde{\psi}_{\alpha_{n}}.
\end{equation}

The generating functional of the 1PI Greens functions in
presence of the bosonic sources $j$ is defined by a Legendre
transform with respect to the fermionic source term $\eta$:
\begin{equation}
\Gamma_{F}[\psi,j]=-W[\eta,j]+\eta_{\alpha}\psi_{\alpha}
\end{equation}
where
\begin{equation}
W=\ln Z[\eta,j], \quad
\psi_{\alpha}=\langle\tilde{\psi}_{\alpha}\rangle =\frac{\partial
W}{\partial\eta_{\alpha}}.
\end{equation}
The fermionic effective action $\Gamma_{F}$ can also be obtained by the implicite functional
integral\footnote{Note that in this formula $\tilde{\psi}$ is
shifted such that $\langle\tilde{\psi}\rangle=0$ and
$\eta_{\alpha}=-\frac{\partial\Gamma_{F}}{\partial\psi_{\alpha}}$
depends on $\psi$.}:
\begin{eqnarray}
\Gamma_{F}[\psi,j]&=&-\ln\int{\mathcal{D}}\tilde{\psi}\exp(\eta_{\alpha}\tilde{\psi}_{\alpha}-S_{j}[\tilde{\psi}+\psi]),
\\\nonumber
S_{j}[\tilde{\psi}]&=&-\frac{1}{2}j_{\alpha \beta}
\tilde{\psi}_{\alpha}\tilde{\psi}_{\beta}+S_{\text{int}}[\tilde{\psi}].
\end{eqnarray}
This form is especially useful to derive the SDE. Taking a
derivative with respect to $\psi$ one finds
\begin{eqnarray}
\frac{\partial\Gamma_{F}}{\partial\psi_{\beta}}\!\!\!\!&&\!\!\!\!=-j_{\beta\alpha_{2}}\psi_{\alpha_{2}}
+\sum_{n}\frac{\lambda^{(n)}_{\beta\alpha_{2}\ldots\alpha_{n}}}{F_{n}}\psi_{\alpha_{2}}
\\\nonumber
\!\!\!\!&\times&\!\!\!\bigg\{
(\Gamma^{(2)}_{F})^{-1}_{\alpha_{3}\alpha_{4}}\cdots(\Gamma^{(2)}_{F})^{-1}_{\alpha_{n-1}\alpha_{n}}
+Z_{\alpha_{3}\ldots\alpha_{n}}+{\mathcal{O}}(\psi^{2})\bigg\}
\end{eqnarray}
where
\begin{equation}
F_{n}=(n-2)(n-4)\cdots 2
\end{equation}
and $\Gamma^{(2)}_{F}$ denotes the second functional derivative.
Here $Z_{\alpha_{3}\ldots\alpha_{n}}$ summarizes all terms containing third and higher
derivatives of $\Gamma$. (The corresponding diagrams involve at least two
vertices.) Taking another derivative with respect to
$\psi_{\alpha}$ and evaluating at $\psi=0$ we find the SDE:
\begin{eqnarray}
\label{equ::sdefermion}
(\Gamma^{(2)}_{F})_{\alpha\beta}&=&-j_{\alpha \beta}
+\sum_{n}\frac{\lambda^{(n)}_{\alpha\beta\alpha_{3}\ldots\alpha_{n}}}{F_{n}}
\\\nonumber
&\times&\bigg\{
(\Gamma^{(2)}_{F})^{-1}_{\alpha_{3}\alpha_{4}}\cdots(\Gamma^{(2)}_{F})^{-1}_{\alpha_{n-1}\alpha_{n}}
+Z_{\alpha_{3}\ldots\alpha_{n}}\bigg\}.
\end{eqnarray}
In this paper we are only interested in the lowest order.
Therefore, from now on, we neglect $Z$, i.e. terms with more than one vertex.

The ``Bosonic Effective Action'' (BEA) \cite{Wetterich:2002ky},
is defined by a different Legendre transform with respect to $j$ for $\eta=0$:
\begin{eqnarray}
\Gamma_{B}[G]&=&-W[0,j]+jG,
\\
\label{equ::grelation} G_{\alpha\beta}&=&\frac{\partial
W}{\partial
j_{\alpha\beta}}=(\Gamma^{(2)}_{F})^{-1}_{\alpha\beta}, \quad
\frac{\partial \Gamma_{B}}{\partial
G_{\alpha\beta}}=j_{\alpha\beta}.
\end{eqnarray}
Since $\Gamma_{F}$ is an even functional of $\psi$ the BEA
contains the same information as $\Gamma_{F}$. Indeed it is
related to $\Gamma_{F}$ by means of functional differential
equations \cite{Wetterich:2002ky} like Eq. \eqref{equ::grelation}. Using this relation we
can conveniently write the SDE \eqref{equ::sdefermion} as
\begin{eqnarray}
\label{equ::sde} G^{-1}_{\alpha\beta}=-j_{\alpha \beta}
+\sum_{n}\frac{\lambda^{(n)}_{\alpha\beta\alpha_{3}\ldots\alpha_{n}}}{F_{n}}
G_{\alpha_{3}\alpha_{4}}\cdots G_{\alpha_{n-1}\alpha_{n}}.
\end{eqnarray}
Eq. \eqref{equ::grelation} then yields a differential equation for
$\Gamma_{B}$
\begin{equation}
\frac{\partial\Gamma_{B}}{\partial G_{\alpha\beta}}=-G^{-1}_{\alpha\beta}
+\sum_{n}\frac{\lambda^{(n)}_{\alpha\beta\alpha_{3}\ldots\alpha_{n}}}{F_{n}}
G_{\alpha_{3}\alpha_{4}}\cdots G_{\alpha_{n-1}\alpha_{n}}.
\end{equation}
By integration\footnote{Note that in our notation
$\frac{\partial G_{\alpha\beta}}{\partial
G_{\gamma\delta}}=\delta_{\alpha\gamma}\delta_{\beta\delta}
-\delta_{\alpha\delta}\delta_{\beta\gamma}$.} one finally finds
\begin{equation}
\label{equ::1vertex} \Gamma_{B}=\frac{1}{2}\Tr\ln
G+\sum_{n}\frac{\lambda^{(n)}_{\alpha_{1}\ldots\alpha_{n}}}{nF_{n}}
G_{\alpha_{1}\alpha_{2}}\cdots G_{\alpha_{n-1}\alpha_{n}},
\end{equation}
the BEA at ``one-vertex order''.
Actually it
is sometimes convenient to introduce an auxiliary effective
action
\begin{equation}
\Gamma_{j}[G,j]=\Gamma_{B}-\frac{1}{2}j_{\alpha\beta}G_{\alpha\beta}
\end{equation}
such that the physical propagator corresponds to the minimum of
$\Gamma_{j}$ (cf. Eq. \eqref{equ::grelation}). The functional
$\Gamma_{j}$ will play the role of a suitable free energy (see \cite{Wetterich:2002ky}
for details).
\section{BEA for Local Interactions}\label{sec::1vertex}
In the following we want to consider local interactions. For
clarity we now write $x$ (or momentum $p$) explicitly and use
latin letters for the remaining indices. The standard procedure
would be the insertion of the ansatz
$G^{-1}_{ab}(x,y)=-j_{ab}(x,y)+\Delta_{ab}(x)\delta(x-y)$ into Eq.
\eqref{equ::sde}. This would yield the SDE for the local gap $\Delta$.
Since the BEA \eqref{equ::1vertex} is related to the SDE
\eqref{equ::sde} by differentiation with respect to $G$ it is not
clear, however, that an effective action functional depending on $\Delta$
can be obtained by integration with respect to $\Delta$.
In presence of several possible gaps this would require suitable ``integrability conditions''
for the system of gap equations. This difficulty can be avoided if we
follow the construction presented in
\cite{Wetterich:2002ky} and start directly from the approximate BEA
\eqref{equ::1vertex}. With
\begin{equation}
\label{equ::localg} g_{ab}(x)=G_{ab}(x,x)
\end{equation}
we have
\begin{eqnarray}
\label{equ::localbea}
\Gamma_{j}&=&\frac{1}{2}\Tr\ln G+\frac{1}{2}\Tr (G j)
\\\nonumber
&&+\int_{x}\sum_{n}\frac{\lambda^{(n)}_{a_{1}\ldots
a_{n}}}{nF_{n}} g_{a_{1}a_{2}}(x)\cdots g_{a_{n-1}a_{n}}(x).
\end{eqnarray}
For this relation it is essential that the interaction is strictly
local. Furthermore, we can use the locality of the interaction in order to
write Eq. \eqref{equ::sde} in the form of a local gap equation
\begin{eqnarray}
\label{equ::ansatz}
G^{-1}_{ab}(x,y)&=&-j_{ab}(x,y)+\Delta_{ab}(x)\delta(x-y).
\end{eqnarray}
We will evaluate the functional $\Gamma_{j}[G]$ for $G_{\alpha\beta}$
taking values corresponding to Eq. \eqref{equ::ansatz}. This is actually a
restriction to a subspace of all possible $G$. However, locality
tells us that the extremum (solution of the SDE) is contained in
this subspace.

Using $j=-G^{-1}+\Delta$ we find (up to a shift in the irrelevant
constant and using $\Delta_{ab}(x,y)=\Delta_{ab}(x)\delta(x-y)$)
\begin{eqnarray}
\label{equ::vorstufe} \nonumber
\Gamma_{j}[g,\Delta]\!\!&=&\!\!-\frac{1}{2}\Tr\ln (-j+\Delta)
-\frac{1}{2}\int_{x}\Delta_{ab}(x)g_{ab}(x)
\\&&\!\!\!\!\!\!\!
+\!\int_{x}\sum_{n}\frac{\lambda^{(n)}_{a_{1}\ldots
a_{n}}}{nF_{n}}g_{a_{1}a_{2}}(x)\cdots g_{a_{n-1}a_{n}}(x).
\end{eqnarray}
For the search of extrema of $\Gamma_{j}$ it is actually
convenient to treat $\Delta$ and $g$ as independent variables. The
extremum of $\Gamma_{j}[g,\Delta]$ then obeys
\begin{equation}
\label{equ::extremum}
\frac{\partial\Gamma_{j}[g,\Delta]}{\partial\Delta}=0,\quad
\frac{\partial\Gamma_{j}[g,\Delta]}{\partial g}=0.
\end{equation}
Evaluating the derivative with respect to $\Delta$ we recover the
inverse of Eq. \eqref{equ::ansatz} for $x=y$,
\begin{equation}
g_{ab}(x)=(-j+\Delta)^{-1}_{ab}(x,x)=g[\Delta(x)].
\end{equation}
Inserting this functional relation into Eq. \eqref{equ::sde} leads
to a gap equation for $\Delta$. In case of a six-fermion
interaction this takes, however, the form of a two-loop equation.

For $n$-fermion interactions with $n>4$ it is more appropriate to go
the other way around and first take a derivative with respect to
$g$. We obtain
\begin{eqnarray}
\nonumber
\Delta_{ab}(x)&=&\sum_{n}\frac{\lambda^{(n)}_{aba_{3}\ldots
a_{n}}}{F_{n}}g_{a_{3}a_{4}}(x)\cdots g_{a_{n-1}a_{n}}(x)
\\
\label{equ::equation} &=&\Delta_{ab}[g(x)],
\end{eqnarray}
which is precisely the value of the gap in Eq. \eqref{equ::sde}.
Inserting $\Delta[g]$ into \eqref{equ::vorstufe} we find the
effective action depending on $g$
\begin{eqnarray}
\label{equ::main} \nonumber \Gamma_{j}[g]&=&-\frac{1}{2}\Tr\ln
(-j+\Delta[g]) -\frac{1}{2}\int_{x}\Delta_{ab}[g](x)g_{ab}(x)
\\
&&+\int_{x}\sum_{n}\frac{\lambda^{(n)}_{a_{1}\ldots
a_{n}}}{nF_{n}}g_{a_{1}a_{2}}(x)\cdots g_{a_{n-1}a_{n}}(x).
\end{eqnarray}
Searching for an extremum yields
\begin{equation}
\label{equ::mot} \frac{\partial{\Gamma_{j}[g]}}{\partial
g}=\left\{(-j+\Delta[g])^{-1} -g\right\}\frac{d\Delta[g]}{d g}=0.
\end{equation}
For $\frac{d\Delta}{dg}\neq 0$ Eq. \eqref{equ::mot} indeed
corresponds to the SDE \eqref{equ::sde}, i.e.
\begin{equation}
\label{equ::gapeq}
g_{ab}(x)=(-j+\Delta[g])^{-1}_{ab}(x,x).
\end{equation}
Eq. \eqref{equ::gapeq} will be our central gap equation. We should point out that
possible extrema of $\Gamma_{j}[g]$ corresponding to
$\frac{d\Delta}{dg}=0$ are not solutions of the gap equation
\eqref{equ::sde} and should be discarded. Finally, we also have
\begin{eqnarray}
\frac{d \Gamma_{j}[g]}{d
g}&=&\frac{d\Gamma_{j}[g,\Delta[g]]}{d g}
\\\nonumber
&=&\frac{\partial\Gamma_{j}[g,\Delta[g]]}{\partial g}+\frac{\partial
\Gamma_{j}[g,\Delta[g]]}{\partial\Delta} \frac{d \Delta[g]}{d g}
\\\nonumber
&=&\frac{\partial \Gamma_{j}[g,\Delta[g]]}{\partial\Delta}
\frac{d \Delta[g]}{d g}.
\end{eqnarray}
Only as long as $\frac{d\Delta[g]}{d g}\neq0$ is fulfilled we can
conclude that a solution of Eq. \eqref{equ::mot} fulfills both
extremum conditions \eqref{equ::extremum}.

The procedure proposed here is quite powerful if $\Tr\ln(-j+\Delta)$ can be
explicitly evaluated as a functional of $\Delta$. Then
$\Gamma_{j}[g]$ allows not only a search for the extremum
(discarding those with $\frac{d\Delta}{dg}=0$) but also a simple
direct comparison of the relative free energy of different local
extrema. This is crucial for the determination of the ground state
in the case of several ``competing gaps''.

Furthermore, the formula \eqref{equ::main} for $\Gamma_{j}$ (and the
corresponding field or gap equation) is now a one loop expression.
This ``one-loop'' form of the equation of motion and the effective action
is very close to what we
would expect from MFT (cf. also the next section). In contrast to
the standard SDE, which is only an equation of motion, we can use Eq.
\eqref{equ::main} to compare the values for the effective action
at different solutions of the equation of motion \eqref{equ::mot},
providing us with information about the stability of a given state.

The one vertex approximation \eqref{equ::equation}, \eqref{equ::main} to
the bosonic effective action is the central tool of this paper
\begin{equation}
\label{equ::v}
\Gamma_{j}[g]=\int_{x}V(g(x))-\frac{1}{2}\int\frac{d^{4}q}{(2\pi)^{4}}\textrm{tr}\ln(-j+\Delta[g]),
\end{equation}
\begin{equation}
\label{equ::classical}
V(g(x))=-\sum_{n}\frac{(n-2)\lambda^{(n)}_{a_{1}\ldots
a_{n}}}{2nF_{n}}g_{a_{1}a_{2}}(x)\cdots g_{a_{n-1}a_{n}}(x).
\end{equation}
The first term may be called ``classical part'' and we have written the second
``one loop term'' as a momentum integral with $\textrm{tr}$ over internal indices
and $-j+\Delta$ involving the Fourier transform of the gap \eqref{equ::equation}
(see \cite{Wetterich:2002ky} for details). As mentioned already we can use $\Gamma_{j}$
in order to find and compare the ``local minima'' of the free energy.
Some care is needed, however, for the computation of susceptibilities or effective masses. Indeed,
we have to be careful when considering
Eq. \eqref{equ::main} at points which are not solutions of Eq.
\eqref{equ::mot}. Going step by step through the procedure above,
we find that if we are not at a solution of \eqref{equ::mot} we do
not necessarily fulfill the ansatz \eqref{equ::ansatz}. Therefore,
at these points we are mathematically not allowed to insert the
ansatz into Eq. \eqref{equ::1vertex}. So, strictly speaking
Eq. \eqref{equ::main} only gives the value of the effective action at
the solution of the equation of motion\footnote{An alternative
would be to choose the gap $\Delta$ as the ``bosonic field''. Inserting Eq. \eqref{equ::ansatz}
into Eq. \eqref{equ::1vertex} we could calculate a functional $\Gamma[\Delta]$.
However, as one can check there are two drawbacks. First, even for four-fermion
interactions, $\Gamma[\Delta]$ is usually unbounded from below when considering
$\Delta\rightarrow\infty$. Second, in the case of a large four-fermion coupling the
``stable'' solution of the field equation is usually a local maximum.}. This is already
much more than what we get from the standard SDE. Going beyond this we would also like to interpret
Eq. \eqref{equ::main} as a reasonable approximation in a small
neighborhood of the solution to the equation of motion.
Remembering $g(x)=\langle\tilde{\psi}(x)\tilde{\psi}(x)\rangle$ it
is suggestive to interpret $g$ as a bosonic field. Eq.
\eqref{equ::equation} gives the (non-linear) ``Yukawa coupling'' of
$g$ to the fermions, i.e. the relation between the gap and the bosonic field.
Thus the term $\sim\Tr\ln$ is the contribution from the
fermionic loop in a background field $g$. The second term in Eq. \eqref{equ::v} can
then be interpreted as the cost in free energy for different
background fields $g$. This interpretation allows us to use
\eqref{equ::main} to calculate the mass and the couplings of the
bosonic field $g$ approximately.
\section{Comparison with MFT}\label{sec::mft}
We refer here to MFT as used in most computations and obtained by partial bosonization
(for a more detailed description of the procedure and its ambiguities see, e.g. \cite{doktor}).
Using the identity
\begin{equation}
\label{equ::mftunity}
\mathbf{1}={\mathcal{N}}\int {\mathcal{D}}\phi\exp(m(-\phi)^{\frac{n}{2}})
={\mathcal{N}}\int {\mathcal{D}}\phi \exp(m(\tilde{\psi}\tilde{\psi}-\phi)^{\frac{n}{2}})
\end{equation}
we can introduce bosonic fields corresponding to fermionic bilinears
$\phi_{ab}(x)=\langle\tilde{\psi}_{a}(x)\tilde{\psi}_{b}(x)\rangle$.
The coefficient $m$ will be chosen such that the insertion of
Eq. \eqref{equ::mftunity} into the functional integral
\eqref{equ::partitionfunction} cancels the purely fermionic interaction
$S_{\textrm{int}}[\tilde{\psi}]$.
We have not displayed the indices in Eq. \eqref{equ::mftunity}. Usually one
restricts the choice to one or several particular index pairs $(a,b)$ (or
linear combinations thereof).
This corresponds to the freedom in the choice of the mean fields $\phi_{ab}(x)$.
The partially bosonized form of Eq. \eqref{equ::iniaction} becomes
\begin{eqnarray}
\nonumber
&&\!\!\!\!\!\!\!\!\!\!\!\!S_{\text{int}}[\phi,\tilde{\psi}]
=\int_{x}\sum_{n}(-1)^{\frac{n}{2}+1}m^{(n)}_{a_{1}b_{1}\ldots a_{\frac{n}{2}}b_{\frac{n}{2}}}
\\\nonumber
&\times&\!\!\!\bigg\{\phi_{a_{1}b_{1}}(x)
\cdots\phi_{a_{\frac{n}{2}}b_{\frac{n}{2}}}(x)
\\\nonumber
&&-\bigg[\tilde{\psi}_{a_{1}}(x)\tilde{\psi}_{b_{1}}(x)\phi_{a_{2}b_{2}}(x)
\cdots\phi_{a_{\frac{n}{2}}b_{\frac{n}{2}}}(x)
\\\nonumber
&&\quad\quad\quad\quad+(1\leftrightarrow 2)+(1\leftrightarrow 3)+\cdots+(1\leftrightarrow\frac{n}{2})\bigg]
\\
&&\quad\quad+\cdots+{\mathcal{O}(\tilde{\psi}^{n-2})}
\bigg\}
\end{eqnarray}
and the functional integral for the partition function \eqref{equ::partitionfunction} has now
to be performed over bosonic as well as fermionic variables.
We note that the coefficients $m$ are only partly determined by $\lambda$, i.e.
\begin{equation}
\label{equ::cond}
m^{(n)}_{a_{1}\ldots a_{n}}=\frac{\lambda^{(n)}_{a_{1}\ldots a_{n}}}{n!}
+S_{a_{1}\ldots a_{n}}.
\end{equation}
Here $S$ is a sum of terms which are symmetric in at least one pair of indices.
The condition \eqref{equ::cond} ensures that the partially bosonized Lagrangian is equivalent
to the original fermionic one. Nevertheless, due to the anticommuting nature of the fermionic variables
$S$ gives a vanishing contribution to the purely fermionic part of the action and is therefore not fixed.

Neglecting the terms ${\mathcal{O}}(\tilde{\psi}^{4})$ and performing the functional integral
over the fermions provides us with the MF-effective action:
\begin{eqnarray}
\label{equ::mft}
\Gamma^{\text{MF}}[\phi]&=&\int_{x}V(\phi(x))-\frac{1}{2}\Tr\ln(-j+\Delta[\phi]),
\\\nonumber
V(\phi)&=&\sum_{n}(-1)^{\frac{n}{2}+1}m^{(n)}_{a_{1}\ldots b_{\frac{n}{2}}}\phi_{a_{1}b_{1}}
\cdots\phi_{a_{\frac{n}{2}}b_{\frac{n}{2}}},
\\\nonumber
\Delta[\phi]_{a_{1}b_{1}}&=&(-1)^{\frac{n}{2}}
\\\nonumber
&&\!\!\!\!\!\!\!\!\!\!\!\!\!\sum_{n}\left[m^{(n)}_{a_{1}\ldots b_{\frac{n}{2}}}\phi_{a_{2}b_{2}}
\cdots\phi_{a_{\frac{n}{2}}b_{\frac{n}{2}}}+\cdots+(1\leftrightarrow\frac{n}{2}) \right].
\end{eqnarray}
This form is strikingly similar to Eq. \eqref{equ::v}. However, the coefficients
in the ``classical potential'' as well as in the ``gap'' $\Delta$ differ.
Furthermore, we note that the coefficients $m$ are not
completely fixed due to the presence of the arbitrary symmetric part $S$ in Eq. \eqref{equ::cond}.
Results can depend on the choice of $S$ (see e.g. \cite{Baier:2000yc,Jaeckel:2003kf}).
This is the so called ''Fierz ambiguity'' because
the addition of $S$ corresponds to a (generalized) Fierz transformation in the fermionic language.
Of course, further considerations as e.g. the stability of the initial bosonic potential
might reduce the freedom in the choice of $S$ somewhat. But, as the example of \cite{Baier:2000yc} shows,
such restrictions are sometimes not even strong enough to get qualitatively the same
phase diagram for
all reasonable choices of $S$.

Eqs. \eqref{equ::vorstufe}, \eqref{equ::main}, \eqref{equ::mot} do not suffer from such an ambiguity since in the
derivation of the SDE \eqref{equ::sdefermion} the coefficient becomes antisymmetrized and
symmetric terms drop out. In \cite{Jaeckel:2003kf} it was shown that the inclusion of
appropriate diagrams for the bosonic fluctuations cures the Fierz ambiguity for four fermion interactions and leads to
the SD-result.
We believe that this holds also for higher fermion interactions.
Nevertheless, the inclusion of the bosonic fluctuations needs a substantial effort.
We therefore propose \eqref{equ::main} as a natural replacement of
Eq. \eqref{equ::mft} which amounts to an ``optimal'' choice of $m$ for
many purposes. Moreover, allowing not only for constant but also for
spatially varying $g$
we can calculate the wave function renormalizations and masses of the bosons. Again, the
BEA gives unambiguous results.

Finally, let us stress again the intuitive argument for the closeness of the two approaches
and explicitly for the similarity of Eqs. \eqref{equ::main} and \eqref{equ::mft}:
we have $g(x)=\langle \tilde{\psi}(x)\tilde{\psi}(x)\rangle$ which is exactly
what one has in mind as a ''mean field''.

\section{Chiral Symmetry Breaking from a Three-Flavor Instanton Interaction} \label{sec::instanton1}
In this section we want to use the method described above to study
chiral symmetry breaking in an NJL-type model with a six-fermion
interaction. We consider the QCD-instanton interaction with three
colors and three flavors
\cite{Shuryak:1982hk,Shifman:uw,Callan:1977gz,'tHooft:fv,Schafer:1996wv} in the
pointlike limit.
The three flavor instanton vertex can be written in the following
convenient form \cite{Wetterich:2000ky}
\begin{eqnarray}
\label{equ::instinteraction}
\nonumber
S_{\text{inst}}[\psi]\!\!&=&
-\frac{\zeta}{6}\int_{x}\epsilon_{a_{1}a_{2}a_{3}}\epsilon_{b_{1}b_{2}b_{3}}
\\\nonumber
&&\quad\quad\bigg\{
(\bar{\psi}^{a_{1}}_{L}\psi^{b_{1}}_{R})
(\bar{\psi}^{a_{2}}_{L}\psi^{b_{2}}_{R})
(\bar{\psi}^{a_{3}}_{L}\psi^{b_{3}}_{R})
\\\nonumber
&&\quad\quad-\frac{1}{8}
(\bar{\psi}^{a_{1}}_{L}\lambda^{z}\psi^{b_{1}}_{R})
(\bar{\psi}^{a_{2}}_{L}\lambda^{z}\psi^{b_{2}}_{R})
(\bar{\psi}^{a_{3}}_{L}\psi^{b_{3}}_{R})
\\\nonumber
&&\quad\quad-\frac{1}{8}
(\bar{\psi}^{a_{1}}_{L}\lambda^{z}\psi^{b_{1}}_{R})
(\bar{\psi}^{a_{2}}_{L}\psi^{b_{2}}_{R})
(\bar{\psi}^{a_{3}}_{L}\lambda^{z}\psi^{b_{3}}_{R})
\\\nonumber
&&\quad\quad-\frac{1}{8}
(\bar{\psi}^{a_{1}}_{L}\psi^{b_{1}}_{R})
(\bar{\psi}^{a_{2}}_{L}\lambda^{z}\psi^{b_{2}}_{R})
(\bar{\psi}^{a_{3}}_{L}\lambda^{z}\psi^{b_{3}}_{R})\bigg]
\\
&&\quad\quad-(R\leftrightarrow L)\bigg\}.
\end{eqnarray}
Here $\lambda^{z}$ are the Gell-Mann matrices with color indices acting as generators of
the $SU(3)_{\text{c}}$ color group. The brackets $(\quad)$
indicate contractions over color and spinor indices.
Within QCD the coupling constant $\zeta$ should be be calculated in terms of the
running gauge coupling. However, already the one loop approximation involves an IR divergent integral over
the instanton size. Therefore, one needs to provide a physical
cutoff mechanism. To avoid this difficulty we treat $\zeta$ as a
free parameter in an effective theory with ultraviolet cutoff $\Lambda$.
Inspection of \eqref{equ::instinteraction} tells us that this
interaction is $U(1)_{A}$-anomalous with a residual
$Z_{3}$-symmetry. This is important because we cannot
restrict ourselves to real condensates from the start.

In order to extract the coupling $\lambda^{(6)}$ we have to antisymmetrize over
flavor indices $(a=1\ldots3)$, color indices $(i=1\ldots3)$
Weyl spinor indices $(\alpha=1,2)$, chirality indices
$(\chi=1,2=L,R)$ and the indices distinguishing between $\psi$ and
$\bar{\psi}$ $(s=1,2)$:
\begin{eqnarray}
\lambda_{m_{1}\ldots m_{6}}&=&P\{\tilde{\lambda}_{m_{1}\ldots m_{6}}\},
\\\nonumber
\tilde{\lambda}_{m_{1}\ldots m_{6}}&=&
\frac{\zeta}{12}\epsilon_{a_{1}a_{2}a_{3}}\epsilon_{a_{4}a_{5}a_{6}}
\delta_{\alpha_{1}\alpha_{4}}\delta_{\alpha_{2}\alpha_{5}}\delta_{\alpha_{3}\alpha_{6}}
\\\nonumber
\times\!\!&[&\!\! \delta_{i_{1}i_{4}}\delta_{i_{2}i_{5}}\delta_{i_{3}i_{6}}
-\frac{1}{8}\lambda^{z}_{i_{1}i_{4}}\lambda^{z}_{i_{2}i_{5}}\delta_{i_{3}i_{6}}
\\\nonumber
&&-\frac{1}{8}\lambda^{z}_{i_{1}i_{4}}\delta_{i_{2}i_{5}}\lambda^{z}_{i_{3}i_{6}}
-\frac{1}{8}\delta_{i_{1}i_{4}}\lambda^{z}_{i_{2}i_{5}}\lambda^{z}_{i_{3}i_{6}}]
\\\nonumber
\times\!\!&[&\!\!\delta_{\chi_{1}1}\delta_{\chi_{2}1}\delta_{\chi_{3}1}
\delta_{\chi_{4}2}\delta_{\chi_{5}2}\delta_{\chi_{6}2}
\\\nonumber
&&-\delta_{\chi_{1}2}\delta_{\chi_{2}2}\delta_{\chi_{3}2}
\delta_{\chi_{4}1}\delta_{\chi_{5}1}\delta_{\chi_{6}1}]
\\\nonumber
\times\!\!&[&\!\!
\delta_{s_{1}2}\delta_{s_{2}2}\delta_{s_{3}2}\delta_{s_{4}1}\delta_{s_{5}1}\delta_{s_{6}1}
\\\nonumber
&&+\delta_{s_{1}1}\delta_{s_{2}1}\delta_{s_{3}1}\delta_{s_{4}2}\delta_{s_{5}2}\delta_{s_{6}2}].
\end{eqnarray}
Here $P$ denotes the sum over all $6!$ permutations of the
multiindices $m_{j}=(a_{j},i_{j},\alpha_{j},\chi_{j},s_{j})$,
$j=1\ldots 6$, with minus signs appropriate for total
antisymmetrization.

As a first example we consider a flavor singlet, color singlet
scalar chiral bilinear
($\sigma=\frac{1}{3}\bar{\psi}^{a}_{L}\psi^{a}_{R}$,
\mbox{$\sigma^{\star}=-\frac{1}{3}\bar{\psi}^{a}_{R}\psi^{a}_{L}$})
\begin{eqnarray}
g_{mn}&=&g_{ai\alpha\chi s,bj\beta\tau t}=\frac{1}{6}\delta_{ab}\delta_{ij}\delta_{\alpha\beta}
\\\nonumber
&\times&\!\!
\bigg[\sigma(\delta_{\chi1}\delta_{\tau2}\delta_{s2}\delta_{t1}-\delta_{\chi2}\delta_{\tau1}\delta_{s1}\delta_{t2})
\\\nonumber
&&\,\,-\sigma^{\star}(\delta_{\chi2}\delta_{\tau1}\delta_{s2}\delta_{t1}
-\delta_{\chi1}\delta_{\tau 2}\delta_{s1}\delta_{t2})\bigg]
\end{eqnarray}
We evaluate
\begin{eqnarray}
\label{equ::delta}
\Delta[g]_{mn}&=&{-}\frac{\lambda_{mm_{2}m_{3}nm_{5}m_{6}}}{8}g_{m_{2}m_{5}}g_{m_{3}m_{6}}
\\\nonumber
&&\!\!\!\!\!\!\!\!\!\!\!\!\!\!\!\!\!\!\!\!\!\!\!\!\!\!\!\!\!\!
={-}6\bigg[\tilde{\lambda}_{mm_{2}m_{3}nm_{5}m_{6}}-\tilde{\lambda}_{mm_{2}m_{3}m_{5}nm_{6}}
+\tilde{\lambda}_{mm_{2}m_{3}m_{5}m_{6}n}\bigg]
\\\nonumber
&\times&\!\!\bigg[g_{m_{2}m_{5}}g_{m_{3}m_{6}}-g_{m_{2}m_{6}}g_{m_{3}m_{5}}\bigg]
\end{eqnarray}
where we have used the fact that $\tilde{\lambda}$ is symmetric
under permutations of the three $\bar{\psi}\psi$ bilinears.
Exploiting the flavor, spin and color structure for
\eqref{equ::delta} expresses the gap in terms of $\sigma$
\begin{eqnarray}
\nonumber
\Delta[g]_{mn}&=&-\frac{{10}}{9}\zeta\delta_{ab}\delta_{ij}\delta_{\alpha\beta}
\bigg[\sigma^{2}(\delta_{\chi1}\delta_{\tau2}\delta_{s2}\delta_{t1}
-\delta_{\chi2}\delta_{\tau1}\delta_{s1}\delta_{t2})
\\
&&\quad\quad-\sigma^{\star
2}(\delta_{\chi2}\delta_{\tau1}\delta_{s2}\delta_{t1}
-\delta_{\chi 1}\delta_{\tau 2}\delta_{s1}\delta_{\tau2})\bigg].
\end{eqnarray}
For the classical potential one obtains
\begin{eqnarray}
\label{equ::class}
V(\sigma)=\frac{20}{9}\zeta(\sigma^{3}+\sigma^{\star 3}).
\end{eqnarray}
For a background $\sigma$ which is constant in space we obtain
the effective potential $U$ ($x=q^2$)
\begin{eqnarray}
\nonumber \label{equ::potentialsigma}
U(\sigma)&=&-\frac{9}{8\pi^2}\int dx \,\,x
[\ln(x+|M_{q}|^2)] +V(\sigma),
\\
M_{q}&=&\frac{10}{9}\zeta \sigma^2+m_{q}.
\end{eqnarray}
Here $m_{q}$ is a current quark mass which we take to be
equal for all quarks. The integral \eqref{equ::potentialsigma}
is, of course, divergent. Our UV regularization simply cuts
it off with $x<\Lambda^2$. Measuring all quantities in units of
$\Lambda$ we can put $\Lambda=1$.
\subsection{The chiral limit $m_{q}=0$}
Let us now search for
extrema of $U(\sigma)$\footnote{Since
$\frac{d\Delta[\sigma]}{d\sigma}\neq 0$ for all $\sigma\neq 0$ we
do not need to worry for non-trivial solutions to be spurious. In
addition $\sigma=0$ is always a solution in the chiral limit.}.
Inspection of $U(\sigma)$ tells us that it is invariant
under the combined operation $\zeta\rightarrow-\zeta$,
$\sigma\rightarrow-\sigma$. This allows us to restrict our
analysis to positive $\zeta$.
In the chiral limit it is useful to parametrize
\begin{equation}
\sigma=|\sigma|\exp(i\alpha).
\end{equation}
From this one finds
\begin{equation}
U(|\sigma|,\alpha)=\frac{40}{9}\zeta|\sigma|^{3}\cos(3\alpha)+f(|\sigma|),
\end{equation}
where $f$ is a function determined by the integral in Eq.
\eqref{equ::potentialsigma}. We can see that the only $\alpha$-dependence
comes from $\cos(3\alpha)$ which is the explicit manifestation of the
$Z_{3}$-symmetry.
It is clear that extrema can only occur at
$\alpha=\frac{n\pi}{3}$, $n\in {\mathbb{Z}}$. Using the
$Z_{3}$-symmetry we can restrict ourselves to
$\alpha=0,\pi$ or restrict ourselves simply
to real $\sigma$.

\begin{figure}[t]
\begin{center}
\scalebox{0.8}[0.8]{
\begin{picture}(350,185)
\Text(66,145)[c]{\scalebox{1.5}[1.5]{$U(\sigma)$}}
\Text(252,30)[c]{\scalebox{1.5}[1.5]{$\sigma$}}
\includegraphics{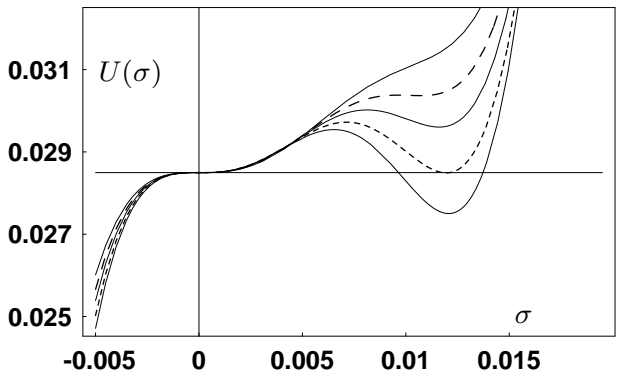}
\end{picture}
}
\end{center}
\caption{Effective potential for various values of the coupling constant increasing from the topmost line
$\zeta=3000$ to the lowest line $\zeta=4200$. The second line (long dashed), for
$\zeta_{\textrm{S}}\approx 3350$, corresponds to the critical coupling for the onset of non-vanishing solutions.
The third line is for $\zeta=3600$ while the fourth (short dashed) denotes the onset
of SSB at $\zeta_{\text{SSB}}\approx 3900$. The horizontal line indicates the value
of $U(\sigma)$ at the trivial solution $\sigma=0$.} \label{fig::beapotential}
\end{figure}

\begin{figure}[t]
\begin{center}
\scalebox{0.8}[0.8]{
\begin{picture}(350,185)
\SetOffset(3,0)
\Text(60,160)[c]{\scalebox{1.8}[1.8]{$|M_{\scalebox{0.5}[0.5]{$q$}}|$}}
\Text(122,40)[c]{\scalebox{1.8}[1.8]{$\zeta_{\scalebox{0.5}[0.5]{\text{S}}}$}}
\Text(208,40)[c]{\scalebox{1.8}[1.8]{$\zeta_{\scalebox{0.5}[0.5]{\text{SSB}}}$}}
\Text(155,53)[c]{\scalebox{1.8}[1.8]{$\tilde{\zeta}_{\scalebox{0.5}[0.5]{\text{SSB}}}$}}
\Text(260,40)[c]{\scalebox{1.8}[1.8]{$\zeta$}}
\put(124.5,32.5) {\vector(1,-1){10}}
\put(198,32.5)  {\vector(-1,-1){10}}
\put(153,42.5) {\vector(1,-2){10}}
\includegraphics{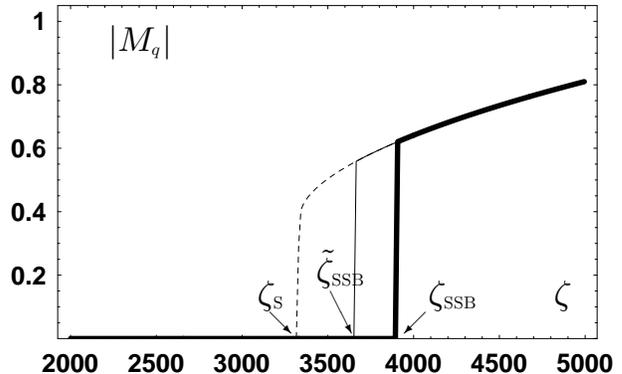}
\end{picture}
}
\end{center}
\caption{Constituent quark mass $|M_{q}|$ (gap) versus the strength of
the six-fermion interaction $\zeta$ in the chiral limit
$m_{q}=0$. The thick line corresponds to the solution with
smallest action. The dashed line reflects the largest non-trivial
solution $\sigma=\sigma_{2}$. Finally, the thin line is obtained by minimizing a
``pseudo effective potential'' $\tilde{U}(\Delta)$ obtained by direct
integration of the SDE with respect to the gap $\Delta$. We find
three special couplings, $\zeta_{\textrm{S}}$ for the onset of
non-trivial solutions to the SDE, $\zeta_{\text{SSB}}$ for the
onset of SSB, i.e. a non-trivial solution that has lower action than
the trivial solution and $\tilde{\zeta}_{\text{SSB}}$ where the
lowest extremum of $\tilde{U}(\Delta)$ becomes non-trivial.
In our approximation we obtain a
first order phase transition. We also see that the ``true'' phase transition
at $\zeta_{\textrm{SSB}}$ differs from the value obtained by a naive integration
of the SDE.} \label{fig::firstorder}
\end{figure}

Taking all this into account we find up to three solutions (cf. Fig. \ref{fig::beapotential}).
By symmetry $\sigma=0$ is a solution for all values of the
coupling. Going to larger couplings we encounter a point
$\zeta_{\textrm{S}}$ where we have two solutions. For even larger
couplings there are three solutions
$0=\sigma_{0}<\sigma_{1}\leq\sigma_{2}$. We know that
$U(\sigma_{1})>U(\sigma_{0}=0)$ therefore
$\sigma_{1}$ is not the stable solution. As can be seen from Figs.
\ref{fig::beapotential}, \ref{fig::firstorder} there is a
range $\zeta_{\textrm{S}}\leq\zeta\leq\zeta_{\text{SSB}}$ where there
exist non-trivial solutions to the SDE but no spontaneous chiral symmetry breaking occurs
because $U(\sigma_{2})\geq U(0)$. The state with $\sigma\neq 0$ is metastable.
This holds as long as $\zeta_{\textrm{S}}<\zeta<\zeta_{\textrm{SSB}}$, with $\zeta_{\textrm{S}}$ the point
of ``spinodal decomposition''. A first order phase transition towards a state with
nonvanishing chiral condensate $\sigma$ occurs as $\zeta$ is increased beyond $\zeta_{\textrm{SSB}}$.
Here the critical value $\zeta_{\textrm{SSB}}$ for the onset of spontaneous chiral symmetry breaking
denotes the value of $\zeta$ for which the two local minima become degenerate, $U(\sigma_{2})=U(0)$.
We point out that in order to calculate $\zeta_{\text{SSB}}$ we
need to know the value of $U$, i.e. information beyond
the SDE.

In Fig. \ref{fig::firstorder} we have plotted the mass gap versus
the six-fermion coupling strength.
We observe the first order
phase transition. This may be expected generically due to the ``cubic term'' in the
classical potential \eqref{equ::class}.

Finally, we would like to remark that in general the free energy does not follow from
a naive integration of the SDE with respect to the gap $\Delta$.
We may call the result of a naive integration of the gap equation the
``pseudo potential'' $\tilde{U}(\Delta)$, or, more generally, the
``pseudo-free energy'' $\tilde{\Gamma}[\Delta]$. In general
$\tilde{\Gamma}[\Delta]$ is \emph{not}
equal to the the BEA $\Gamma_{j}[G]$, not even at solutions of
the SDE. Indeed, as can be seen from Fig. \ref{fig::firstorder},
the results for physical quantities like the effective fermion
mass can differ. The underlying reason for this is that the gap
$\Delta$ is not the correct integration variable. The SDE is
obtained by a $G$-derivative of the BEA functional
$\Gamma_{j}[G]$. Therefore, in order to reconstruct
$\Gamma_{j}[G]$, we have to integrate with respect to $G$. As
can be seen from Eq. \eqref{equ::equation} $\Delta$ is, in
general, not even a linear function of $G$. Simple integration
with respect to $\Delta$ therefore neglects the Jacobi matrix,
which is a non-trivial function of $\Delta$ for interactions more
complicated than a four-fermion interaction.

\subsection{Non-vanishing current quark masses $m_{q}\neq 0$}
The non-vanishing current quark mass explicitly breaks the
residual $Z_{3}$-symmetry. The effective potential
$U(|\sigma|,\alpha)$ does no longer depend on
$\cos(3\alpha)$ only, and we have to look at the complete complex
$\sigma$-plane for possible extrema.
Moreover, for $m_{q}=0$, $U(\sigma)$ is completely
symmetric under $\zeta\rightarrow -\zeta, \sigma\rightarrow
-\sigma$. Therefore, we could restrict ourselves to $\zeta\geq 0$.
For $m_{q}\neq 0$ we need to add the transformation
$m_{q}\rightarrow -m_{q}$. We can still restrict ourselves to
positive $\zeta$ but we need to consider both positive and
negative $m_{q}$.

In the case of $m_{q}\neq 0$ we still encounter an extremum of
$U(\sigma)$ at $\sigma=0$. However, in this case it is
not a solution of the SDE. It is a spurious solution due to
$\frac{d\Delta[\sigma]}{d\sigma}=0$. The difference to the chiral
limit is that the derivative of the effective potential
$U(\sigma)$ now has only a simple zero while in the chiral
limit it has a twofold zero. After dividing the field equation by
$\frac{d\Delta[\sigma]}{d\sigma}$ a simple zero remains only in the chiral limit,
giving a solution of the SDE.

\begin{figure}[t]
\begin{center}
\scalebox{0.8}[0.8]{
\begin{picture}(350,185)
\Text(70,70)[c]{\scalebox{1.8}[1.8]{$|\delta M_{\scalebox{0.5}[0.5]{$q$}}|$}}
\Text(70,140)[c]{\scalebox{1.8}[1.8]{$\zeta_{\scalebox{0.5}[0.5]{\text{SSB}}}$}}
\Text(200,32)[c]{\scalebox{1.8}[1.8]{$m_{\scalebox{0.5}[0.5]{$q$}}$}}
\includegraphics{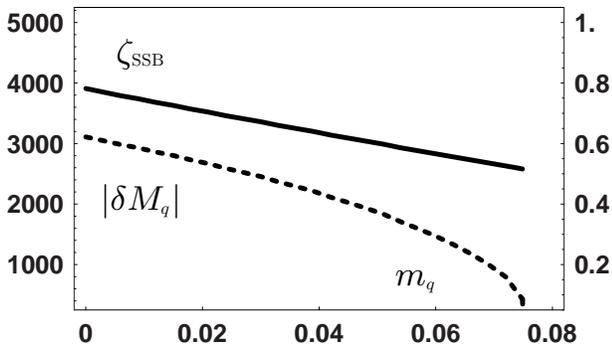}
\end{picture}
}
\end{center}
\caption{Dependence of the critical coupling $\zeta_{\textrm{SSB}}$ on the current quark
mass $m_{q}$ (solid line) and jump in the fermion mass at the
phase transition $|\delta M_{q}|$ (dashed line, scale on the right hand side).
We observe that there exists
a critical $m_{q}\approx 0.076$ above which the
phase transition is replaced by a continuous crossover.} \label{fig::currentquark}
\end{figure}
Although chiral symmetry is now broken explicitly we can still
observe a first-order phase transition signaled by a jump in the
fermion mass. The critical coupling for the phase transition depends on
$m_{q}$ as depicted in Fig. \ref{fig::currentquark}. The
critical line ends at
$m_{q}=m_{q,\textrm{crit}}\approx 0.076$.
For $m_{q}>m_{q,\textrm{crit}}$ we have no first
order phase transition in our approximation but rather
find a continuous crossover. We have also plotted
in Fig. \ref{fig::currentquark} the jump in the constituent quark mass
$|\delta M_{q}|$ between the phases with low and large $|\sigma|$ at
the critical coupling $\zeta_{\textrm{SSB}}$. It vanishes at the end of
the critical line at $m_{q,\textrm{crit}}$.

\section{Color-Octet Condensation} \label{sec::instanton2}
In the preceeding section we have considered only one direction in the
space of all possible $g$ resulting in a phase diagram for chiral
symmetry breaking. Let us now consider the more general case where
we also allow for a non-vanishing expectation value in the
color-octet channel, more explicitly in the color-flavor locked
direction
\begin{eqnarray}
g_{mn}&=&g_{ai\alpha\chi s,bj\beta\tau t}
\\\nonumber
&&=\frac{1}{6}\delta_{ab}\delta_{ij}\delta_{\alpha\beta}
\bigg[\sigma(\delta_{\chi1}\delta_{\tau2}\delta_{s2}\delta_{t1}-\delta_{\chi2}\delta_{\tau1}\delta_{s1}\delta_{t2})
\\\nonumber
&&\quad\quad\,\,-\sigma^{\star}(\delta_{\chi2}\delta_{\tau1}\delta_{s2}\delta_{t1}
-\delta_{\chi1}\delta_{\tau 2}\delta_{s1}\delta_{t2})\bigg]
\\\nonumber
&&+\frac{1}{4\sqrt{6}}\lambda^{z}_{ab}\lambda^{z}_{ij}\delta_{\alpha\beta}
\bigg[\chi(\delta_{\chi1}\delta_{\tau2}\delta_{s2}\delta_{t1}-\delta_{\chi2}\delta_{\tau1}\delta_{s1}\delta_{t2})
\\\nonumber
&&\quad\quad\,\,-\chi^{\star}(\delta_{\chi2}\delta_{\tau1}\delta_{s2}\delta_{t1}
-\delta_{\chi1}\delta_{\tau 2}\delta_{s1}\delta_{t2})\bigg].
\end{eqnarray}
A condensate in this direction would yield a very interesting phenomenology
for QCD where the gluons acquire a mass via the Higgs mechanism
and are associated with the $\rho$-mesons \cite{Wetterich:1999vd,Wetterich:2000pp}.
A pointlike six-quark interaction seems too simple in this case for a realistic model of QCD
-- it does not account for the important infrared cutoff of the instanton size
due to the effective gluon mass \cite{Wetterich:2000ky}.
Nevertheless, we find a study of a pointlike interaction an interesting preparation
for the treatment of a more realistic model.

Following the outline of the previous section we obtain the gap
\begin{eqnarray}
\nonumber
\Delta[g]_{mn}&=&-\zeta\bigg[\frac{5}{864}(192\sigma^{2}+2\sqrt{6}\sigma\chi-7\chi^{2})\delta_{ab}\delta_{ij}
\\\nonumber
&&\quad-\frac{5}{288}\chi(2\sqrt{6}\sigma+\chi)\delta_{ai}\delta_{bj}\bigg]
\\\nonumber
&&\quad\times\delta_{\alpha\beta}
\bigg[\delta_{\chi1}\delta_{\tau2}\delta_{s2}\delta_{t1}
-\delta_{\chi2}\delta_{\tau1}\delta_{s1}\delta_{t2}\bigg]
\\
&&\quad\quad+(s\leftrightarrow t,\sigma\rightarrow\sigma^{\star},\chi\rightarrow\chi^{\star})
\end{eqnarray}
and the classical potential
\begin{eqnarray}
\label{equ::classoctet}
V(\sigma,\chi)=\zeta\bigg[\frac{20}{9}\sigma^{3}-\frac{5}{18}\sigma\chi^{2}-\frac{5}{54\sqrt{6}}\chi^{3}+c.c.\bigg].
\end{eqnarray}
This results in the effective potential
\begin{eqnarray}
\label{equ::octetpotential}
U(\sigma,\chi)&=&V[\sigma,\chi]
\\\nonumber
&-&\frac{1}{8\pi^2}\int_{x} dx\,\,x[8\ln(x+|M_{8}|^{2})
+\ln(x+|M_{1}|^{2})]
\end{eqnarray}
where the singlet and octet masses for the ``constituent quarks'' associated
with the low mass baryons \cite{Wetterich:1999vd,Wetterich:2000pp} are given by
\begin{eqnarray}
\label{equ::masses}
M_{1}&=&\frac{5}{54}\zeta(12\sigma^{2}-\sqrt{6}\sigma\chi-\chi^{2})+m_{q}
\\
M_{8}&=&\frac{5}{864}\zeta(192\sigma^{2}+2\sqrt{6}\sigma\chi-7\chi^{2})+m_{q}.
\end{eqnarray}

We start by a search of extrema of $U$ for real $\sigma$ and $\chi$.
In the chiral limit every point on the line $\chi=-2\sqrt{6}\sigma$
($M_{1}=0$ and $M_{8}=0$) has the same value of
$U=0$, and both derivatives with respect to $\sigma$ and
$\chi$ vanish. However, this is one of the spurious solutions
mentioned in Sect. \ref{sec::1vertex} where $\frac{\partial\Delta}{\partial g}$
vanishes. (In our case $\Delta=(M_{1},M_{8})$ and $g=(\sigma,\chi)$
are two component vectors and $\frac{\partial \Delta}{\partial g}$ stands
for the Jacobian.) Direct insertion into Eq. \eqref{equ::gapeq} ($\Delta(g)=0$)
shows that
on this line only the point $(\sigma,\chi)=(0,0)$ is a true
solution to the SDE.
Restricting both $\sigma$ and $\chi$ to be real we have not found a
solution of the gap equation with $\chi\neq 0$.
Thus, we have not identified a solution for which
the condensate would break color symmetry but not parity.

For the most general case of complex $\sigma$ and $\chi$ things
are considerably more difficult since we now have to search for an
extremum of a potential which depends on four real parameters. We
checked several values of the coupling constant. So far we have not
found a solution which has a lower free energy than the minimum of the
free energy for vanishing octet condensate $\chi=0$.

Still, we would like to point out that the potential is unbounded
from below in various directions, including those with $\chi\neq 0$.
Therefore, a physical cutoff mechanism
like the one discussed in \cite{Wetterich:2000ky} or a different approximation
to the ``classical action'' which makes the potential bounded from
below may provide additional solutions.
In this context we stress that the instanton interaction \eqref{equ::instinteraction}
should not be confounded
with realistic QCD. The gluon fluctuations have been omitted here. Perhaps even
more important, the effective gluon mass for nonvanishing $\chi$ should lead to
an effective $\chi$-dependence of the coupling $\zeta$ \cite{Wetterich:2000ky}.
Also the flat direction\footnote{This flat direction is also present for $m_{q}\neq 0$.} of $U$
for $\chi=-2\sqrt{6}\sigma$ remains intriguing. So
far we have not yet understood the reason why $U$, $M_{1}$ and $M_{8}$ all vanish
simultaneously on this ``line''. A small additional contribution
to $U$ (for example from the gauge boson fluctuations) could
lift this degeneracy and lead to a minimum of $U$ somewhat away from this line, such
that $\frac{\partial\Delta}{\partial g}$ does not vanish anymore.

Finally, let us compare the result of our SD calculation with the MFT result
analogous to the computation in \cite{Wetterich:1999sh}. However,
we use here the corrected instanton vertex of \cite{Wetterich:2000ky}
(without the cutoff mechanism for the instanton interaction considered there)
and no other interaction. We apply the formulae of sect. \ref{sec::mft} even though
in our case the integral \eqref{equ::mftunity} is not well defined.
Adopting the same normalization for the MF as for the condensates $g$ in the SD calculation
and taking the mean field as suggested by the brackets in
\eqref{equ::instinteraction} without further Fierz transformation the
``classical MF potential'' is
\begin{equation}
V(\sigma_{\textrm{\tiny{MF}}},\chi_{\textrm{\tiny{MF}}})
=-\zeta(\sigma^3_{\textrm{\tiny{MF}}}+\frac{1}{6}\sigma_{\textrm{\tiny{MF}}}\chi^{2}_{\textrm{\tiny{MF}}})
+c.c..
\end{equation}
For the ``one-loop'' Potential $U_{\textrm{\tiny{MF}}}$ we recover
the form \eqref{equ::octetpotential} but with
masses
\begin{eqnarray}
M^{\textrm{\tiny{MF}}}_{1}&=&-\zeta(\sigma^{2}_{\textrm{\tiny{MF}}}
+\frac{1}{3}\sqrt{\frac{2}{3}}\sigma_{\textrm{\tiny{MF}}}\chi_{\textrm{\tiny{MF}}}
+\frac{1}{18}\chi^{2}_{\textrm{\tiny{MF}}})
\\\nonumber
M^{\textrm{\tiny{MF}}}_{8}&=&-\zeta(\sigma^2_{\textrm{\tiny{MF}}}
-\frac{1}{12\sqrt{6}}\sigma_{\textrm{\tiny{MF}}}\chi_{\textrm{\tiny{MF}}}
+\frac{1}{18}\chi^{2}_{\textrm{\tiny{MF}}}).
\end{eqnarray}
This quite different from our SD result Eqs. \eqref{equ::classoctet}, \eqref{equ::masses}.
In particular
the $\chi^{3}$-term in the classical potential is absent in the MFT calculation. Moreover,
the sign between the $\sigma^{3}$-term and the $\sigma\chi^{2}$ is different, too. This
demonstrates that the difference is more than an overall normalization of the potential or the
fields. On the one side this highlights once more the importance of the Fierz ambiguity.
On the other side it is not obvious if a suitable mean field formulation exists
at all which would reproduce the results of the SDE. In view of all the problems
of MFT we would like to argue that the SDE is clearly superior for our problem.

\section{Summary and Conclusions}\label{sec::conclusions}
Integrating the lowest order Schwinger-Dyson equation (SDE) for a
multifermion-interaction we obtain the bosonic effective action at
''one-vertex''-level. Within this approximation we find an ''one-loop''
 expression for the
SDE even in case of interactions involving more than four fermions.
Although this gap equation is formally very similar to mean field theory, it does
not suffer from the ambiguities of the latter arising from the bosonization procedure.
We also propose a simple one loop formula for the free energy functional at the extrema
which can be used in order to compare different local extrema.

We apply our method to a six-fermion interaction
resembling the instanton induced quark vertex for three colors and flavors.
We compute the solutions of the gap equation and the minimum of the free energy
in dependence of the coupling strength $\zeta$.
For small current quark masses $m_{q}$ the gap equation has several solutions,
corresponding to different local extrema of the free energy.
The free energy of the different extrema is compared by use of the one loop formula
for the bosonic effective action. We find a first order transition to a phase
with chiral symmetry breaking as the coupling of the six-quark vertex $\zeta$
increases beyond a critical value $\zeta_{\textrm{SSB}}$. Thereby
the value of $\zeta_{\textrm{SSB}}$ is larger than the value $\zeta_{\textrm{S}}$,
the minimal coupling for which solutions with a non-vanishing chiral order
parameter exist. (Note that $\zeta_{\textrm{SSB}}$ equals $\zeta_{\textrm{S}}$ only
in the case of the perhaps more familiar second order transition.)
The critical line in the space of the current quark mass $m_{q}$ and $\zeta$ ends
at a mass $m_{q,\textrm{crit}}\approx 0.076$
(in units of the UV cutoff near $1\textrm{GeV}$). For $m_{q}>m_{q,\textrm{crit}}$
the phase diagram is characterized by a continuous crossover.
We also investigate possible color octet condensates in the color-flavor-locked direction.
In the approximation of a pointlike instanton induced six-quark interaction no
phase with non-vanishing color octet condensate is visible. In this respect
our work should be considered as a starting point for a more realistic instanton
induced interaction, where the dependence of the instanton solution on the
value of the octet condensate is taken into account.

We conclude that the bosonic effective action provides a practical tool for the understanding
of fermionic systems where interactions involving more
than four fermions play an important role. Beyond the gap equation it provides for a free energy.
In contrast to mean field theory the lowest order is well defined and gives an unambiguous answer.
\begin{acknowledgements}
The authors would like to thank J. Berges and H. Gies for useful discussions.
J.J.~acknowledges financial support by the Deutsche
Forschungsgemeinschaft under contract Gi 328/1-2.
\end{acknowledgements}

\end{document}